\documentclass[aps,prb,showpacs,twocolumn]{revtex4-2}
\usepackage{amsfonts}
\usepackage{amssymb}
\usepackage{amsmath}
\usepackage{graphicx}
\usepackage{epsfig}
\usepackage{subfigure}
\usepackage{appendix}
\usepackage{hyperref}
\usepackage{color}

\hypersetup{hypertex=true,
colorlinks=true,
linkcolor=blue,
urlcolor = red,
citecolor=red}

\begin{document}

\title{Topologically protected two-fluid edge states}
\author{K. L. Zhang}
\author{Z. Song}
\email{songtc@nankai.edu.cn}
\affiliation{School of Physics, Nankai University, Tianjin 300071, China}
\date{\today}

\begin{abstract}
Edge states reveal the nontrivial topology of energy band in the bulk. As
localized states at boundaries, many-body edge states may obey a special
symmetry that is broken in the bulk. When local particle-particle
interaction is induced, they may support a particular property. We consider
an extended two-dimensional Su-Schrieffer-Heeger Hubbard model and examine
the appearance of $\eta$-pairing states, which are excited eigenstates
related to superconductivity. In the absence of Hubbard interaction, the
energy band is characterized by topologically invariant polarization in
association with edge states. In the presence of on-site Hubbard
interaction, $\eta$-pairing edge states appear in the topologically
nontrivial phase, resulting in the condensation of pairs at the boundary. In
addition, as Hamiltonian eigenstates, the edge states contain paired
fermions and unpaired fermions. Neither affects the other; they act as
two-fluid states. From numerical simulations of many-body scattering
processes, a clear manifestation and experimental detection scheme of
topologically protected two-fluid edge states are provided.
\end{abstract}

\maketitle

\section{Introduction}

\label{Introduction}

Topology in matter permits the existence of edge states, which have a strong
immunity to distortions of the underlying architecture. Although they are
believed to be inherited from the topology of the bulk, these edge states
present some properties only around the boundary and are forbidden in the
bulk. Topological insulator in condensed matter physics is a well-known
example of this phenomenon. Although such a system is an insulator in the
bulk, it permits electron conductance along the edges, resulting in
quantized Hall conductance \cite{klitzing1980new, laughlin1981quantized,
thouless1982quantized, niu1985quantized}. In general, the strong interaction
between fermions can affect the topology of a fermion system and break the
bulk-boundary correspondence (BBC) \cite{chiu2016classification}.
Nevertheless, recent work \cite{zhang2021quantum} shows that quantum spin
system also exhibits BBC even in a thermal state. This indicates that
although the BBC reveals a nontrivial topology of the energy band, it has a
particular feature in the presence of interaction.

Beyond the realm of the well-known topological insulator, recent studies on
the twisted bilayer graphene (TBG) show that at a series of magic twist
angles, the moir\'{e} pattern yields the flat low-energy bands that promise
strong electronic correlations \cite{bistritzer2011moire, wu2018hubbard,
wong2020cascade}. The interplay of lattice geometry and many-body
interactions induces exotic quantum states including superconducting \cite%
{xu2018topological, peri2021fragile} and correlated insulating \cite%
{xie2020nature} behaviors, which have been achieved in experiments \cite%
{cao2018unconventional, cao2018correlated}. An interesting question is
whether such exotic quantum states exist in a topological system, where the
energy band of edge states plays the same role as the moir\'{e} flat band,
in the presence of strong electronic correlations. 
The $\eta$ pairing proposed by Yang \cite{yang1989eta} is a promising
mechanism to address this problem. In the absence of Hubbard interaction $U$%
, an $\eta$ pair has zero energy in a bipartite lattice, thus the flat band
appears when considering multiple $\eta$ pairs formed by 
electrons with momentums $\mathbf{k}$ and $\boldsymbol{\pi}-\mathbf{k}$ \cite%
{yang1989eta}. There are two  
common grounds between the flat bands of TBG and that of many-body edge
states: They are formed by breaking the translational symmetry and have the
potential to enhance electronic correlations. Importantly, such paired states become long lived in the
presence of strong Hubbard repulsion \cite{rosch2008metastable,
strohmaier2010observation, sensarma2010lifetime, hofmann2012doublon}. In
addition, recent theoretical works \cite{kaneko2019photoinduced,
tindall2019heating} suggest that the $\eta$-pairing states can be induced by
pulse irradiation or heating. On the other hand, as a phenomenological
theory, the two-fluid model was proposed several decades ago to explain the
behavior of superfluid helium \cite{tisza1938transport, landau1941theory}
and a conventional superconductor \cite{ginzburg2009theory}. It postulates
that a superconductor possesses two parallel channels, one superconducting
and one normal. Although it is a useful tool, so far two-fluid states have
yet to be investigated in the framework of quantum mechanics, appearing as a
many-body Hamiltonian eigenstate. For instance, even the usual BCS wave
function \cite{bardeen1957theory} is not an eigenstate of a Hamiltonian
system with a local potential energy.

\begin{figure}[t]
\centering
\includegraphics[width=0.5\textwidth]{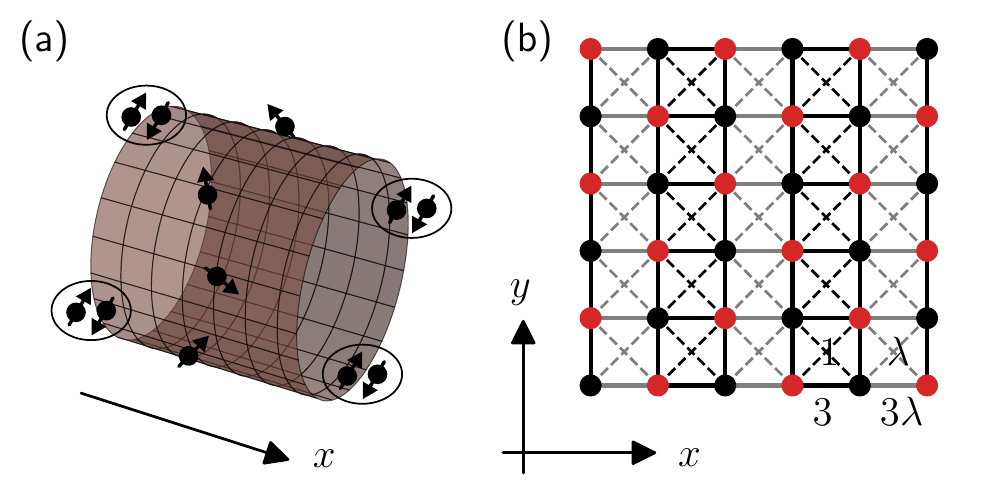}
\caption{(a) Schematic of the lattice with cylindrical boundary condition.
The lattice supports $\protect\eta$-pairing edge states, whereas the bulk
states are unpaired fermions. (b) Details of the lattice, which is an
extended 2D SSH Hubbard model depicted by the Hamiltonian in Eq. (\protect
\ref{H_2D}). The black and red dots represent sublattices A and B,
respectively. The solid and dashed lines represent the nearest-neighbor and
next-nearest-neighbor hopping, respectively.}
\label{fig1}
\end{figure}

In this paper, we consider an extended two-dimensional (2D)
Su-Schrieffer-Heeger (SSH) \cite{su1979solitons} model with on-site Hubbard
interaction $U$. Specifically, we examine the appearance of $\eta $-pairing
edge states [see Fig.~\ref{fig1}(a)]. In the absence of Hubbard interaction,
the energy band is characterized by topologically invariant polarization in
association with edge states. When $U$ switches on, the bulk states do not
support the formation of $\eta $ pairs, which appear only at the edge and
possess an off-diagonal long-range order (ODLRO) \cite{yang1962concept} in
the topologically nontrivial phase. We further propose a concept of
topologically protected two-fluid edge states. As many-body eigenstates of a
2D SSH Hubbard model, the two-fluid edge states contain two components: $%
\eta $-pairing fermions and unpaired fermions, which do not affect each
other. The signature of the two-fluid edge states is observable in the
dynamic behavior of resonant transmission. Based on the non-Hermitian
quantum mechanics \cite{bender1998real, mostafazadeh2002pseudo,
muller2008exceptional, heiss2012physics, rotter2015review}, we also develop
a method to resolve the scattering problem involving multiple interacting
fermions.

The remainder of this paper is organized as follows. In Sec. \ref%
{Model and two-fluid states}, we introduce the concept of two-fluid states
by a Hubbard model. In Sec. \ref{Topological eta pairing edge states}, we
demonstrate the existence of topological $\eta$-paring edge states in an
extended 2D SSH Hubbard model, which also supports two-fluid edge states and
the properties are studied in Sec. \ref{Resonant transmission} through
dynamics of resonant transmission. In Sec. \ref{Summary}, we summarize our
results.

\section{Model and two-fluid states}

\label{Model and two-fluid states}

We begin with the Hamiltonian model of a general form and a brief summary of
related results. The Hamiltonian is in the following form%
\begin{equation}
H=H_{0}+H_{\mathrm{intra}},
\end{equation}%
on two sublattices $\mathrm{A}$ and $\mathrm{B}$. Here $H_{0}$ is the
Hamiltonian of a simple Hubbard model with bipartite lattice symmetry 
\begin{equation}
H_{0}=\sum_{\mathbf{r}\in \mathrm{A},\mathbf{r}^{\prime }\in \mathrm{B}%
}\sum_{\sigma =\uparrow ,\downarrow }t_{\mathbf{rr}^{\prime }}c_{\mathbf{r}%
,\sigma }^{\dagger }c_{\mathbf{r}^{\prime },\sigma }+\text{\textrm{H.c.}}%
+U\sum_{\mathbf{r}\in \mathrm{A}\cup \mathrm{B}}n_{\mathbf{r},\uparrow }n_{%
\mathbf{r},\downarrow },
\end{equation}%
where the operator $c_{\mathbf{r},\sigma }$ ($c_{\mathbf{r},\sigma
}^{\dagger }$) is the usual annihilation (creation) operator of a fermion
with spin $\sigma \in \left\{ \uparrow ,\downarrow \right\} $ at site $%
\mathbf{r}$, and $n_{\mathbf{r},\sigma }=c_{\mathbf{r},\sigma }^{\dagger }c_{%
\mathbf{r},\sigma }$ is the number operator for a particle of spin $\sigma $
on site $\mathbf{r}$. As many previous studies \cite{rosch2008metastable,
strohmaier2010observation, sensarma2010lifetime, hofmann2012doublon,
kaneko2019photoinduced, tindall2019heating}, here the interaction is purely
on-site. Nevertheless, the weak nearest-neighbor interaction does not break
the paired states \cite{lin2014sudden} considered below. Unlike almost all
studies on the $\eta $-pairing states, we consider an extra term $H_{\mathrm{%
intra}}$, representing the intrasublattice hopping. This term breaks the
bipartite symmetry of $H_{0}$.

We first review well-established model properties of $H_{0}$ that are
crucial to our conclusion (see Appendix \ref{A} for more details). First, $%
H_{0}$\ possesses SU(2) symmetry characterized by the generators $%
s^{+}=\left( s^{-}\right) ^{\dagger }=\sum_{\mathbf{r}}s_{\mathbf{r}}^{+}$
and $s^{z}=\sum_{\mathbf{r}}s_{\mathbf{r}}^{z} $, where $s_{\mathbf{r}%
}^{+}=c_{\mathbf{r},\uparrow }^{\dagger }c_{\mathbf{r},\downarrow } $ and $%
s_{\mathbf{r}}^{z}=\left( n_{\mathbf{r},\uparrow }-n_{\mathbf{r},\downarrow
}\right) /2$. Second, one can define the following operators 
\begin{equation}
\eta ^{+} =\left( \eta ^{-}\right) ^{\dagger }=\sum_{\mathbf{r}}\eta _{ 
\mathbf{r}}^{+},\quad \eta ^{z} =\sum_{\mathbf{r}}\eta _{\mathbf{r}}^{z},
\end{equation}
with $\eta _{\mathbf{r}}^{+}=c_{\mathbf{r},\uparrow }^{\dagger }c_{\mathbf{r}
,\downarrow }^{\dag }$ ($-c_{\mathbf{r},\uparrow }^{\dagger }c_{\mathbf{r}
,\downarrow }^{\dag }$)\ for $\mathbf{r}\in A$ ($\mathbf{r}\in B$), and $%
\eta _{\mathbf{r}}^{z}=\left( n_{\mathbf{r},\uparrow }+n_{\mathbf{r}
,\downarrow }-1\right) /2$, satisfying commutation relation $[\eta _{ 
\mathbf{r}}^{+},$ $\eta _{\mathbf{r}}^{-}]=2\eta _{\mathbf{r}}^{z}$ and $%
[\eta _{\mathbf{r}}^{z},$ $\eta _{\mathbf{r}}^{\pm }]=\pm \eta _{\mathbf{r}%
}^{\pm }$. Straightforward algebra shows that%
\begin{equation}
\left[ H_{0}-U\eta ^{z},\eta ^{\pm }\right] =\left[ H_{0},\eta ^{z}\right]
=0,  \label{H_eta_commute}
\end{equation}
which is guaranteed by the bipartite lattice symmetry. These two properties
allow the construction of two types of eigenstates of $H_{0}$: ferromagnetic
(FM) and antiferromagnetic (AFM).

An $m$-fermion FM eigenstate with energy $\sum_{\left\{ \mathbf{k}\right\}
}^{m}\varepsilon _{\mathbf{k}}$ can be expressed as follows:%
\begin{equation}
\left\vert \psi _{\mathrm{FM}}(m,l)\right\rangle =\left( s^{-}\right)
^{\left( m-l\right) /2}\prod_{\left\{ \mathbf{k}\right\} }^{m}c_{\mathbf{k}%
,\uparrow }^{\dagger }\left\vert \mathrm{Vac}\right\rangle ,
\end{equation}%
where $\left\vert \mathrm{Vac}\right\rangle $ is the vacuum state of the
fermion $c_{\mathbf{r},\sigma }$, and $c_{\mathbf{k},\uparrow }^{\dagger } $%
\ is the eigenmode of $H_{0}$ with $U=0$ ---that is, $[c_{\mathbf{k},\sigma
}^{\dagger },H_{0}(U=0)]=-\varepsilon _{\mathbf{k}}c_{\mathbf{k},\sigma
}^{\dagger }$. Eigenstate $\left\vert \psi _{\mathrm{FM}}(m,l)\right\rangle $
is a saturated FM state, given that it is also an eigenstate of $s^{2}$ and $%
s^{z}$, with eigenvalues $m\left( m/2+1\right) /2$ and $l/2,$ respectively.

An $n$-pair AFM eigenstate can be expressed as%
\begin{equation}
\left\vert \psi _{\mathrm{AFM}}(n)\right\rangle =\left( \eta ^{+}\right)
^{n}\left\vert \mathrm{Vac}\right\rangle ,
\end{equation}%
which obeys $H_{0}\left\vert \psi _{\mathrm{AFM}}(n)\right\rangle
=nU\left\vert \psi _{\mathrm{AFM}}(n)\right\rangle $ and $s^{2}\left\vert
\psi _{\mathrm{AFM}}(n)\right\rangle =0$. Obviously, an $\eta $-pairing
state is spin singlet. Unlike all other spin singlet states, an AFM $\eta $%
-pairing eigenstate is independent of the detailed structure of the
bipartite lattice, regardless of whether it contains short-, long-, or even
infinite-range hopping terms. This feature is characterized by a correlator
related to the off-diagonal element of the reduced density matrix of the
system \cite{yang1962concept, yang1989eta}.

Importantly, there is a mixture of two types of states, referred to as
two-fluid state%
\begin{equation}
\left\vert \psi _{\mathrm{2F}}(n,m,l)\right\rangle =\left( \eta ^{+}\right)
^{n}\left( s^{-}\right) ^{\left( m-l\right) /2}\prod_{\left\{ \mathbf{k}%
\right\} }^{m}c_{\mathbf{k},\uparrow }^{\dagger }\left\vert \mathrm{Vac}%
\right\rangle ,
\end{equation}%
which is a common eigenstate of $H_{0}$, $s^{2}$ and $s^{z}$ with
eigenvalues $\sum_{\left\{ \mathbf{k}\right\} }^{m}\varepsilon _{\mathbf{k}%
}+nU$, $m\left( m/2+1\right) /2$ and $l/2,$ respectively. States $\left\vert
\psi _{\mathrm{2F}}(n,m,l)\right\rangle $ are demonstrated to be related to
ODLRO and superconductivity for finite $n/N$ in a large-$N$ limit \cite%
{yang1962concept, yang1990so, singh1991exact}, where $N$ is the total number
of lattice sites. In a comparison of states $\left\vert \psi _{\mathrm{AFM}%
}(n)\right\rangle $ and $\left\vert \psi _{\mathrm{FM}}(m,l)\right\rangle $,
the first is a condensation of pairs, acting as a Bose-Einstein condensate,
whereas the second is a free electron gas, acting as a normal conductor.
Notably, eigenstate $\left\vert \psi _{\mathrm{2F}}(n,m,l)\right\rangle $
contains two components, paired and single electrons, and no scattering
occurs between single electrons and $\eta $ pairs. This mechanism suggests
the presence of resonant transmission channels for a Hubbard cluster in the
state $\left\vert \psi _{\mathrm{AFM}}(n)\right\rangle $\ as a scattering
center. This can be verified by examining the scattering dynamics of an
input Gaussian wave packet with resonant energy. Here we emphasize that the
bipartite symmetry plays an important role in the formation of the AFM $\eta$%
-pairing states and the two-fluid states.

We wish to determine what happens in the presence of $H_{\mathrm{intra}}$.
Obviously, the paired states are suppressed in general. However, the
foregoing analysis remains true if there exists an invariant subspace
spanned by a set of many-body states $\left\{ \left\vert \psi _{\mathrm{p}%
}\right\rangle \right\} $, satisfying%
\begin{equation}
H_{\mathrm{intra}}\left\vert \psi _{\mathrm{p}}\right\rangle =0.
\end{equation}%
We refer to these types of states as conditional $\eta $-pairing states.
Furthermore, it would be interesting when states $\left\{ \left\vert \psi _{%
\mathrm{p}}\right\rangle \right\} $\ have special physical significant
because an $\eta $-pairing state has ODLRO or superconductivity. We will
consider a concrete example in which states $\left\{ \left\vert \psi _{%
\mathrm{p}}\right\rangle \right\} $\ are topologically protected edge
states. These conditional $\eta $-pairing states permit the existence of
two-fluid edge states, which support not only the superconductivity but also
a single-fermion transmission channel at the system boundary.

\section{Topological $\protect\eta $-pairing edge states}

\label{Topological eta pairing edge states}

We consider an extended 2D SSH Hubbard model on an $N_{x}\times N_{y}$
square lattice, as presented in the schematic in Fig.~\ref{fig1}(b). The
Hamiltonian consists of two parts%
\begin{equation}
H_{2\text{\textrm{D}}}=H_{\text{\textrm{0}}}+H_{\mathrm{intra}}  \label{H_2D}
\end{equation}%
with $H_{\text{\textrm{0}}}=3\sum_{\mathbf{r},\sigma }(t_{x}c_{x,y,\sigma
}^{\dag }c_{x+1,y,\sigma }$ $+c_{x,y,\sigma }^{\dag }c_{x,y+1,\sigma }+%
\mathrm{H.c.})+U\sum_{\mathbf{r}}n_{\mathbf{r},\uparrow }n_{\mathbf{r}%
,\downarrow }$ and $H_{\mathrm{intra}}=\sum_{\mathbf{r},\sigma
}t_{x}(c_{x,y,\sigma }^{\dag }c_{x+1,y+1,\sigma }$ $+c_{x,y,\sigma }^{\dag
}c_{x+1,y-1,\sigma })+\mathrm{H.c.}$. Here, $x$ and $y$ [$\mathbf{r}=\left(
x,y\right) $] are the lattice indexes in the $\hat{x}$ and $\hat{y}$
directions, respectively, and the parameter $t_{x}=\lambda ^{\left( x\text{ }%
\mathrm{mod}\text{ }2\right) }$. The hopping term $H_{\mathrm{intra}}$
breaks the bipartite lattice symmetry, as indicated by the dashed lines in
Fig.~\ref{fig1}(b).

We first focus on the interaction-free case with $U=0$, where a single
parameter $\lambda $ controls the topological quantum phase. The system is
in the topologically nontrivial phase within $\left\vert \lambda \right\vert
<1 $ (see Appendix \ref{B}). With the open boundary condition in $\hat{x}$
direction [see Fig.~\ref{fig1}(a)] and in the large-$N_{x}$ limit, the
system supports two degenerate edge modes 
\begin{eqnarray}
|\psi _{k_{y},\sigma }^{\mathrm{L}}\rangle &=&\Omega \sum_{x}\left( -\lambda
\right) ^{x-1}a_{x,k_{y},\sigma }^{\dag }\left\vert \mathrm{Vac}%
\right\rangle ,  \notag \\
|\psi _{k_{y},\sigma }^{\mathrm{R}}\rangle &=&\Omega \sum_{x}\left( -\lambda
\right) ^{N_{x}-x}b_{x,k_{y},\sigma }^{\dag }\left\vert \mathrm{Vac}%
\right\rangle ,  \label{single_edge_state}
\end{eqnarray}%
with the eigenenergy $E^{\mathrm{L}/\mathrm{R}}=6\cos k_{y}$ and the
normalization constant $\Omega =\sqrt{1-\lambda ^{2}}$. Here, $\left(
a_{x,k_{y},\sigma },b_{x,k_{y},\sigma }\right) $ $=N_{y}^{-1/2}\sum_{y}%
\left( c_{2x-1,y,\sigma },c_{2x,y,\sigma }\right) e^{-ik_{y}y}$, and this
transformation does not break the bipartite lattice symmetry. We note that 
\begin{equation}
H_{\mathrm{intra}}|\psi _{k_{y},\sigma }^{\mathrm{L/R}}\rangle =0,
\label{condition}
\end{equation}%
which indicates that edge states $\{|\psi _{k_{y},\sigma }^{\mathrm{L/R}%
}\rangle \}$\ span an invariant subspace with bipartite lattice symmetry and
make possible the formation of $\eta $-pairing eigenstates when $U$ is
switched on. Notably, Eq. (\ref{condition}) still holds when disordered
perturbation is introduced. This indicates that the $\eta $-pairing edge
states are topologically protected. In the trivial phase $\left\vert \lambda
\right\vert >1$, or when periodic boundary condition in both directions are
taken, these paired eigenstates are absent. This is another important
characterization of topological system.

\begin{figure}[t]
\centering
\includegraphics[width=0.5\textwidth]{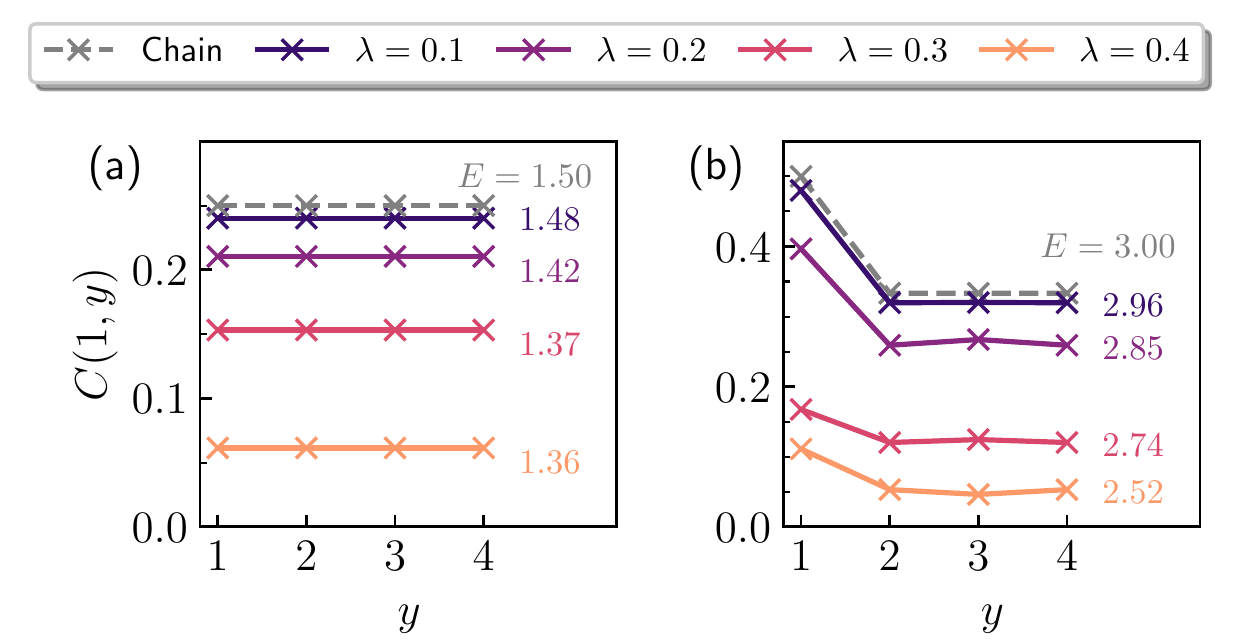}
\caption{Numerical results of the correlator $C\left( 1,y\right) $ defined
in Eq. (\protect\ref{correlator}) for systems with different numbers of
fermions and parameters $\protect\lambda $. The lattice sizes and Hubbard
interaction strengths are all $(N_{x},N_{y})=(5,4)$ and $U=1.5$. The number
of fermions in (a) and (b) is two ($\uparrow \downarrow $) and four ($%
\uparrow\uparrow \downarrow \downarrow $), respectively. For comparison, the
correlators for a bipartite chain are also plotted (dashed lines). The
corresponding energies are indicated to the right of each line.}
\label{fig2}
\end{figure}

To verify the existence of the $\eta $-pairing edge states $|\psi _{\text{%
\textrm{edge}}}\rangle$, we study the system $H_{2\text{\textrm{D}}}$ in
two- and four-fermion subspaces with Hubbard interaction $U$ through exact
diagonalization. To characterize the $\eta $-pairing edge states, we
introduce the following correlator \cite{yang1962concept, yang1989eta}%
\begin{equation}
C\left( y^{\prime },y\right) =\langle \psi _{\text{\textrm{edge}}}|\eta
_{x=1,y^{\prime }}^{\dag }\eta _{x=1,y}|\psi _{\text{\textrm{edge}}}\rangle ,
\label{correlator}
\end{equation}%
where $\eta _{x,y}=\left( -1\right) ^{y}c_{x,y,\downarrow }c_{x,y,\uparrow }$%
. The lattice index $x$ is fixed at the left end in Eq. (\ref{correlator}),
because the lattice possesses inversion symmetry, and for an edge state, the
correlator should vanish in the bulk. For the eigenstate $|\psi _{\text{%
\textrm{edge}}}\rangle $ possessing ODLRO, the correlator $C\left( y^{\prime
},y\right) $ is a constant when $\left\vert y^{\prime }-y\right\vert $
increases. This is a characteristic of superconductivity. When $\lambda
\approx 0$, the presence of $\eta $-pairing edge states is clear; the two
chains at the boundaries, which support the edge states, are isolated from
the bulk of the cylinder, and the boundary chains can be regarded as
approximate bipartite lattices. For the $\eta $-pairing state $\left\vert
\psi (n)\right\rangle =(\eta ^{\dag })^{n}|\mathrm{vac}\rangle $ in a
bipartite lattice, the correlator is \cite{yang1989eta} 
\begin{equation}
C\left( y^{\prime },y\right) =\frac{n\left( N_{y}-n\right) }{N_{y}\left(
N_{y}-1\right) },y^{\prime }\neq y.  \label{correlator_even}
\end{equation}%
This bipartite lattice corresponds to the isolated chain with $N_{y}$ sites
at the boundary.

The $\eta $-pairing edge states are bound pairs localized at the boundary,
which have substantial pairing energy and near-zero kinetic energy; thus,
the pair density at the boundary $\mathcal{N}_{\text{\textrm{edge}}%
}=\sum_{y}\langle \psi _{\text{\textrm{edge}}}|n_{1,y,\uparrow
}n_{1,y,\downarrow }|\psi _{\text{\textrm{edge}}}\rangle $ is useful in the
search for the $\eta $-pairing edge states. One can search for $\eta $%
-pairing edge states in eigenstates with large $\mathcal{N}_{\text{\textrm{%
edge}}}$. We perform the numerical calculation for systems with different
numbers of particles and parameters $\lambda $.

The numerical results of the correlator $C\left( y^{\prime },y\right) $ for
the $\eta $-pairing edge states are shown in Fig. \ref{fig2}. The
corresponding local particle density is presented in Appendix \ref{C}. The
lattice size of the system is set as $(N_{x},N_{y})=(5,4)$, where $N_{x}$ is
truncated to an odd number such that the edge states appear in only one
boundary. The interaction strength is $U=1.5$, and the number of particles
are two (one spin up and one spin down, marked as $\uparrow \downarrow $),
and four ($\uparrow \uparrow \downarrow \downarrow $) for Figs. \ref{fig2}%
(a) and (b), respectively. In the invariant subspaces of two and four
fermions, there exist only one $\eta$-pairing state with one-pairing and
two-pairing, respectively. A higher filling ratio will involve the bulk
fermions, which have no contribution to the ODLRO. The correlators of a
uniform chain in Eq. (\ref{correlator_even}) are plotted as a comparison.
One can see that until $\lambda =0.4$, these two states have the long-range
correlation of $C\left( 1,y\right) $. The comparison between the energies of
two and four fermions indicates that there is no interaction between the
paired fermions in small $\lambda$ limit, which is equivalent to finite $%
\left\vert \lambda \right\vert <1$ with large $N_{x}$. In Appendix \ref{C},
we also give the numerical results of three fermions, and the approximate
results of even- and odd-number fermions in a larger system. In these cases,
the previous conclusions are still valid.

\section{Resonant transmission}

\label{Resonant transmission}

In the preceding section, we demonstrate the existence of the $\eta $%
-pairing edge states from the correlator. A natural question is that whether
system $H_{\mathrm{2D}}$ supports a two-fluid state containing single
fermions and $\eta $ pairs localized at the boundary of the system. We
attempt to answer this question and demonstrate the properties of the
two-fluid edge states.

Particle beam scattering is a conventional technique for detecting the
nature of matter \cite{jin2021symmetry}. Because no scattering occurs
between the single fermions and $\eta $ pairs, resonant transmission
provides a means to present the features of two-fluid edge states. The
detection of resonant transmission for a many-body state is somewhat
challenging, both theoretically and experimentally. Nevertheless,
exceptional point (EP) \cite{muller2008exceptional, heiss2012physics,
rotter2015review} dynamics in non-Hermitian quantum mechanics can reduce the
difficulty of the calculation. For a system with parameters at EP, two or
more eigenvalues along with their associated eigenstates become identical,
leading to unidirectional dynamics. This allows us to employ EP dynamics to
simulate the perfect resonant transmission of particles. It can shorten the
lengths of the input and output leads, and the scattering process in a
Hermitian system can be effectively treated as the dynamics of a
non-Hermitian system (see Appendix \ref{D}). In the following, we present
the resonant transmission dynamics of the two-fluid edge states. For
comparison, the scattering dynamics between a single fermion and an FM edge
state are also examined.

\begin{figure}[tbh]
\centering
\includegraphics[width=0.5\textwidth]{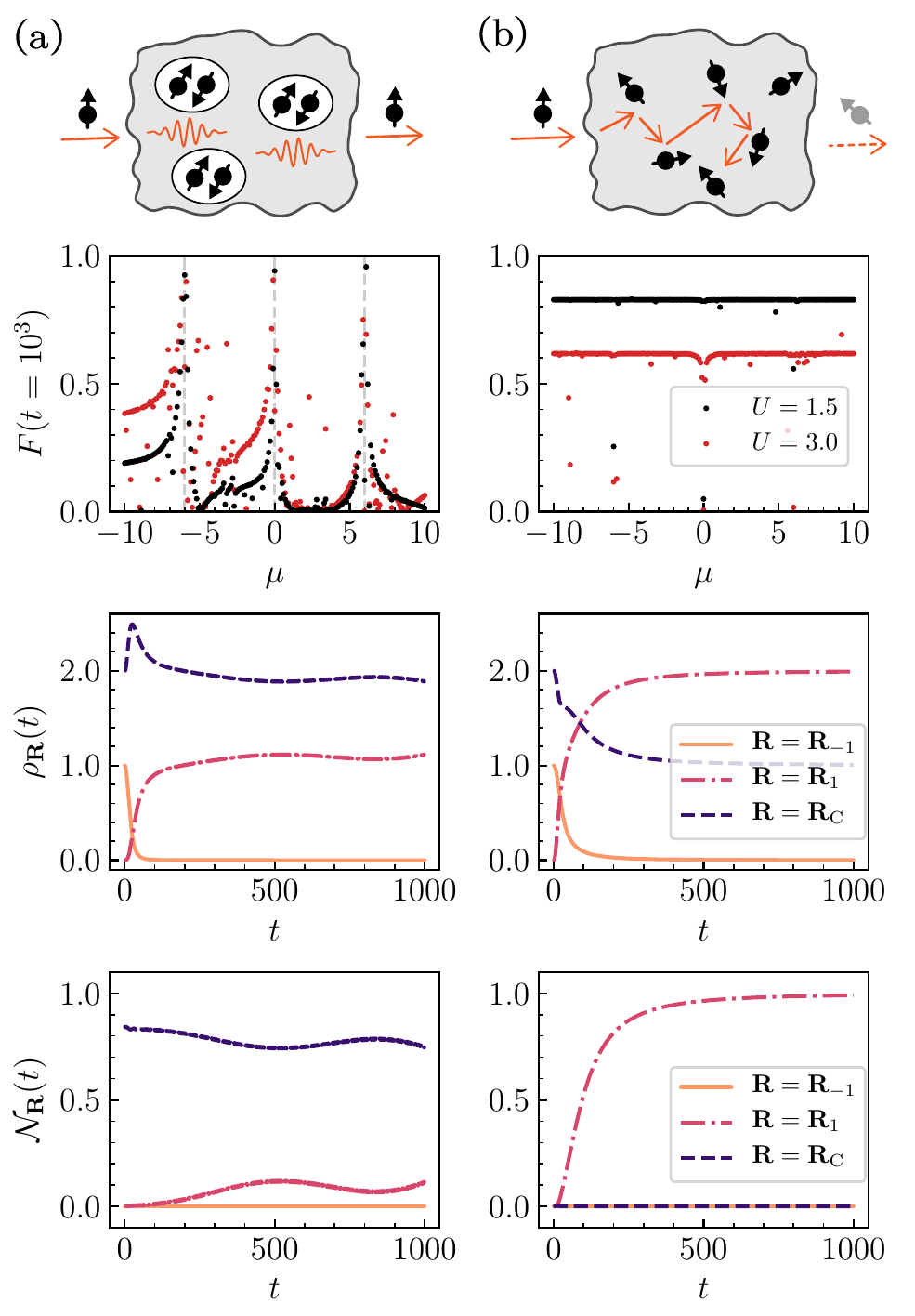}
\caption{(a) Resonant transmission process. (b) Ordinary scattering process.
The top panel is a schematic of the two processes. The bottom panel presents
the corresponding numerical results of fidelity defined in Eq. (\protect\ref%
{fidelity}) at time $t=1000$ as a function of $\protect\mu$ for $U=1.5$ and $%
3$, as well as the particle density $\protect\rho _{\mathbf{R}}(t) $ and the
pair density $\mathcal{N}_{\mathbf{R}}(t)$ defined in Eqs. (\protect\ref%
{particle_density}) and (\protect\ref{particlas_paring}) for parameters $%
\protect\mu =0$ and $U=1.5$. The other parameters are $(N_{x},N_{y})=(5,4)$, 
$\protect\lambda=0.2$ and $J=0.1$. The scales of the Hamiltonian and time $t$
are taken as dimensionless.}
\label{fig3}
\end{figure}

We consider the dynamics in a non-Hermitian system, which consists of the 2D
SSH Hubbard model $H_{2\text{\textrm{D}}}$ and two extra sites (with indexes 
$j=-1$ and $1$), connected by the unidirectional hopping 
\begin{eqnarray}
H_{\text{\textrm{NH}}} &=&H_{2\text{\textrm{D}}}+J\sum_{\sigma =\uparrow
,\downarrow }\left( c_{\alpha ,\sigma }^{\dagger }c_{-1,\sigma }+c_{1,\sigma
}^{\dagger }c_{\beta ,\sigma }\right)  \notag \\
&&+\mu \sum_{\sigma =\uparrow ,\downarrow }\sum_{j=\pm 1}c_{j,\sigma
}^{\dagger }c_{j,\sigma },
\end{eqnarray}%
where $c_{\alpha ,\sigma }^{\dagger }$ and $c_{\beta ,\sigma }$ are fermion
operators at the edge of the cylinder $H_{2\text{\textrm{D}}}$. A schematic
of the system configuration is presented in Appendix \ref{D}. When $\mu =E^{%
\mathrm{L/R}}=6\cos k_{y}$ is considered, a Jordan block with order of three
should appear in single-particle subspace (see Appendix \ref{D}). For the
many-body case, if the two-fluid states remain eigenstates of $H_{2\text{%
\textrm{D}}}$, it still supports the Jordan block with order of three. This
can be verified by examining the dynamic process. We take the initial state
as $\left\vert \psi (0)\right\rangle =c_{-1,\uparrow }^{\dagger }|\psi _{%
\text{\textrm{edge}}}^{\mathrm{a}}\rangle $, with the AFM $\eta $-pairing
edge state $|\psi _{\text{\textrm{edge}}}^{\mathrm{a}}\rangle $ satisfying $%
s^{2}|\psi _{\text{\textrm{edge}}}^{\mathrm{a}}\rangle =0$, and calculate
the time evolution $\left\vert \psi (t)\right\rangle =e^{-iH_{\text{\textrm{%
NH}}}t}\left\vert \psi (0)\right\rangle /|e^{-iH_{\text{\textrm{NH}}%
}t}\left\vert \psi (0)\right\rangle |$. The expected final state is $%
\left\vert \psi _{\mathrm{final}}\right\rangle =c_{1,\uparrow }^{\dagger
}|\psi _{\text{\textrm{edge}}}^{\mathrm{a}}\rangle $ when resonant
transmission occurs. Here, the numerical computations are performed by using
a uniform mesh in conducting time discretization.

To characterize the procedure, we introduce three quantities: the fidelity
between the expected state and the evolved state%
\begin{equation}
F(t)=\left\vert \left\langle \psi _{\mathrm{final}}\right. \left\vert \psi
(t)\right\rangle \right\vert ;  \label{fidelity}
\end{equation}%
the particle density of the evolved state at different spatial regions%
\begin{equation}
\rho _{\mathbf{R}}(t)=\sum_{\mathbf{r=R}}\sum_{\sigma =\uparrow ,\downarrow
}\left\langle \psi (t)\right\vert n_{\mathbf{r},\sigma }\left\vert \psi
(t)\right\rangle ,  \label{particle_density}
\end{equation}%
where $\mathbf{R}=\mathbf{R}_{-1}$ and $\mathbf{R}_{1}$ respectively
represent the extra sites with indexes $j=-1$ and $1$, and $\mathbf{R}=%
\mathbf{R}_{\mathrm{C}}$ represents the sites of the scattering center $H_{2%
\text{\textrm{D}}}$; and the pair density 
\begin{equation}
\mathcal{N}_{\mathbf{R}}(t)=\sum_{\mathbf{r=R}}\left\langle \psi
(t)\right\vert n_{\mathbf{r},\uparrow }n_{\mathbf{r},\downarrow }\left\vert
\psi (t)\right\rangle .  \label{particlas_paring}
\end{equation}%
Figure. \ref{fig3}(a) presents the numerical results of the fidelity after a
sufficiently long time as a function of $\mu $, and the two other quantities
as functions of time $t$. As expected, the peaks appear at $\mu =-6$, $0$
and $6$, where the single fermion is resonantly transmitted from site $%
\mathbf{R}_{-1}$ to site $\mathbf{R}_{1}$. This can also be seen in the
particle density $\rho _{\mathbf{R}}(t)$ and the pair density $\mathcal{N}_{%
\mathbf{R}}(t)$ for $\mu =0$ and $U=1.5$.

For comparison, in Fig. \ref{fig3}(b), we consider the time evolution of
another initial state $\left\vert \psi (0)\right\rangle =c_{-1,\uparrow
}^{\dagger }|\psi _{\text{\textrm{edge}}}^{\mathrm{b}}\rangle $, with the FM
edge state $|\psi _{\text{\textrm{edge}}}^{\mathrm{b}}\rangle $ satisfying$\
s^{2}|\psi _{\text{\textrm{edge}}}^{\mathrm{b}}\rangle =2$, $s_{z}|\psi _{%
\text{\textrm{edge}}}^{\mathrm{b}}\rangle =0$, and the state $\left\vert
\psi _{\mathrm{final}}\right\rangle =c_{1,\uparrow }^{\dagger }|\psi _{\text{%
\textrm{edge}}}^{\mathrm{b}}\rangle $. The numerical results indicate that
the single fermion is scattered by the $\mathrm{FM}$ state $|\psi _{\text{%
\textrm{edge}}}^{\mathrm{b}}\rangle $ at any $\mu $ value due to the Hubbard
interaction. When $\mu=0$ and $U=1.5$, two fermions are transmitted to site $%
\mathbf{R}_{1}$ after a sufficiently long time.

\section{Summary}

\label{Summary}

We present a concept of topologically protected two-fluid edge states and a
means for their detection. We demonstrate the existence of such states in
the 2D SSH Hubbard model, which provides an analog to the topological
insulator. Such a material behaves as a conductor in its interior but
possesses a surface containing superconducting states. In other words, the
condensation of $\eta $ pairs can exist only on the surface of the material.
We also determine that a two-fluid state can be formed as a many-body
eigenstate of a realistic Hubbard model. By employing EP dynamics, a
technique of the numerical simulation is developed to overcome the
computational difficulty in the scattering problem of many-body system,
which involve numerous basis vectors. It is expected to measure the resonant
transmission in experiment via peaks in transmission coefficient \cite%
{landauer1957spatial, nazarov2009quantum}. At present, the possible
experimental implementation to explore the two-fluid edge states is
ultracold fermions in an optical lattice \cite{mazurenko2017cold}.

\acknowledgments This work was supported by National Natural Science
Foundation of China (under Grant No. 11874225).

\renewcommand{\thesection}{}

\section*{Appendix}

In this Appendix, we present \ref{A}. Symmetries, FM and AFM eigenstates for 
$H_{0}$; \ref{B}. Solution of the extended 2D SSH model; \ref{C}. More exact
numerical results and approximate numerical results; and \ref{D}.
Non-Hermitian description of resonant transmission. \renewcommand{%
\thesubsection}{\Alph{subsection}}

\subsection{Symmetries, FM and AFM eigenstates for $H_{0}$}

\label{A} \setcounter{equation}{0} \renewcommand{\theequation}{A%
\arabic{equation}}

Considering the Hubbard Hamiltonian $H_{0}$ in the main text, one can
defined two sets of pseudo-spin operators ($s^{\pm },s^{z}$) and ($\eta
^{\pm },\eta ^{z}$). The first set is 
\begin{eqnarray}
s^{+} &=&\left( s^{-}\right) ^{\dagger }=\sum_{\mathbf{r}}s_{\mathbf{r}}^{+},
\notag \\
s^{z} &=&\sum_{\mathbf{r}}s_{\mathbf{r}}^{z},
\end{eqnarray}%
where the local operators $s_{\mathbf{r}}^{+}=c_{\mathbf{r},\uparrow
}^{\dagger }c_{\mathbf{r},\downarrow }$ and $s_{\mathbf{r}}^{z}=\left( n_{ 
\mathbf{r},\uparrow }-n_{\mathbf{r},\downarrow }\right) /2$ obey the Lie
algebra, i.e., $[s_{\mathbf{r}}^{+},$ $s_{\mathbf{r}}^{-}]=2s_{\mathbf{r}%
}^{z}$, and $[s_{\mathbf{r}}^{z},$ $s_{\mathbf{r}}^{\pm }]=\pm s_{\mathbf{r}%
}^{\pm }$. The second set is%
\begin{eqnarray}
\eta ^{+} &=&\left( \eta ^{-}\right) ^{\dagger }=\sum_{\mathbf{r}}\eta _{ 
\mathbf{r}}^{+},  \notag \\
\eta ^{z} &=&\sum_{\mathbf{r}}\eta _{\mathbf{r}}^{z},
\end{eqnarray}%
with $\eta _{\mathbf{r}}^{+}=c_{\mathbf{r},\uparrow }^{\dagger }c_{\mathbf{r}
,\downarrow }^{\dag }$ ($-c_{\mathbf{r},\uparrow }^{\dagger }c_{\mathbf{r}
,\downarrow }^{\dag }$)\ for $\mathbf{r}\in A$ ($\mathbf{r}\in B$), and $%
\eta _{\mathbf{r}}^{z}=\left( n_{\mathbf{r},\uparrow }+n_{\mathbf{r}
,\downarrow }-1\right) /2$ satisfying commutation relation $[\eta _{\mathbf{r%
}}^{+},$ $\eta _{\mathbf{r}}^{-}]=2\eta _{\mathbf{r}}^{z}$, and $[\eta _{ 
\mathbf{r}}^{z},$ $\eta _{\mathbf{r}}^{\pm }]=\pm \eta _{\mathbf{r}}^{\pm }$%
. Straightforward algebra shows the symmetries of $H_{0}$:%
\begin{equation}
\left[ H_{0},s^{\pm }\right] =\left[ H_{0},s^{z}\right] =0,
\label{H_s_commute}
\end{equation}%
and 
\begin{equation}
\left[ H_{0}-U\eta ^{z},\eta ^{\pm }\right] =\left[ H_{0},\eta ^{z}\right]
=0,  \label{H_eta_commu}
\end{equation}
which will be employed to construct eigenstates of the Hamiltonian.

Starting from the diagonalization form of $H_{0}$\ with zero $U$,%
\begin{equation}
H_{0}(U=0)=\sum_{\mathbf{k},\sigma =\uparrow ,\downarrow }\varepsilon _{ 
\mathbf{k}}c_{\mathbf{k},\sigma }^{\dagger }c_{\mathbf{k},\sigma },
\end{equation}%
one can construct an $m$-fermion FM eigenstate of $H_{0}(U\neq 0)$ 
\begin{equation}
\left\vert \psi _{\mathrm{FM}}(m,m)\right\rangle =\prod_{\left\{ \mathbf{k}
\right\} }^{m}c_{\mathbf{k},\uparrow }^{\dagger }\left\vert \mathrm{Vac}
\right\rangle ,
\end{equation}%
with $\left\vert \mathrm{Vac}\right\rangle $ being the vacuum state of
fermion $c_{\mathbf{r},\sigma }$, since the $U$ term has no effect on the
fermions with aligned spin polarization. The symmetry in Eq. (\ref%
{H_s_commute}) permits the existence of eigenstates%
\begin{equation}
\left\vert \psi _{\mathrm{FM}}(m,l)\right\rangle =\left( s^{-}\right)
^{\left( m-l\right) /2}\prod_{\left\{ \mathbf{k}\right\} }^{m}c_{\mathbf{k}
,\uparrow }^{\dagger }\left\vert \mathrm{Vac}\right\rangle ,
\end{equation}%
with $l=-m,-m+2,...,m-4,m-2,m$, which obeys 
\begin{equation}
H_{0}\left\vert \psi _{\mathrm{FM}}(m,l)\right\rangle =\sum_{\left\{ \mathbf{%
\ k}\right\} }^{m}\varepsilon _{\mathbf{k}}\left\vert \psi _{\mathrm{FM}
}(m,l)\right\rangle .
\end{equation}%
These states are referred to as FM states since they obey%
\begin{equation}
s^{2}\left\vert \psi _{\mathrm{FM}}(m,l)\right\rangle =\frac{m}{2}(\frac{m}{%
2 }+1)\left\vert \psi _{\mathrm{FM}}(m,l)\right\rangle ,  \notag
\end{equation}%
and%
\begin{equation}
s^{z}\left\vert \psi _{\mathrm{FM}}(m,l)\right\rangle =\frac{l}{2}\left\vert
\psi _{\mathrm{FM}}(m,l)\right\rangle .
\end{equation}%
Similarly, one can construct a set of AFM eigenstates based on the symmetry
in Eq. (\ref{H_eta_commu}). An $n$-pair has the form%
\begin{equation}
\left\vert \psi _{\mathrm{AFM}}(n)\right\rangle =\left( \eta ^{+}\right)
^{n}\left\vert \mathrm{Vac}\right\rangle ,
\end{equation}%
which obeys 
\begin{equation}
H_{0}\left\vert \psi _{\mathrm{AFM}}(n)\right\rangle =nU\left\vert \psi _{ 
\mathrm{AFM}}(n)\right\rangle ,  \notag
\end{equation}%
and%
\begin{equation}
s^{2}\left\vert \psi _{\mathrm{AFM}}(n)\right\rangle =0.
\end{equation}%
Obviously, an $\eta $-pairing state is spin singlet. In the main text, a
two-fluid eigenstates is constructed based on the above two types of states.

\subsection{Solution of the extended 2D SSH model}

\label{B} \setcounter{equation}{0} \renewcommand{\theequation}{B%
\arabic{equation}} 

Taking the periodic boundary condition in both directions, the
interaction-free 2D SSH Hamiltonian in $\mathbf{k}$ space can be written as%
\begin{equation}
H_{2\text{\textrm{D}}}(U=0)=\sum_{\mathbf{k}}\sum_{\sigma =\uparrow
,\downarrow }\left( 
\begin{array}{cc}
a_{\mathbf{k,}\sigma }^{\dag } & b_{\mathbf{k,}\sigma }^{\dag }%
\end{array}
\right) h_{\mathbf{k}}\left( 
\begin{array}{c}
a_{\mathbf{k,}\sigma } \\ 
b_{\mathbf{k,}\sigma }%
\end{array}
\right) ,
\end{equation}%
where the core matrix is 
\begin{equation}
h_{\mathbf{k}}=\left( 
\begin{array}{cc}
6\cos k_{y} & \vartheta _{\mathbf{k}} \\ 
\vartheta _{\mathbf{k}}^{\ast } & 6\cos k_{y}%
\end{array}
\right) ,  \notag
\end{equation}%
with $\vartheta _{\mathbf{k}}=\left( \lambda +e^{-ik_{x}}\right) \left(
3+2\cos k_{y}\right) $, based on the Fourier transformation 
\begin{equation}
\left( a_{\mathbf{k,}\sigma },b_{\mathbf{k,}\sigma }\right) =\left(
N_{x}N_{y}\right) ^{-1/2}\sum_{\mathbf{r}}\left( c_{2x-1,y,\sigma
},c_{2x,y,\sigma }\right) e^{-i\mathbf{k\cdot r}}.
\end{equation}%
We note that the system satisfies time reversal symmetry and is invariant
under the inversion symmetry 
\begin{equation}
\mathcal{T}h_{\mathbf{k}}\mathcal{T}^{-1}\mathcal{=}h_{-\mathbf{k}},Rh_{ 
\mathbf{k}}R^{-1}=h_{-\mathbf{k}},
\end{equation}%
where $\mathcal{T}=K$ is conjugation operator and $R=\left( 
\begin{array}{cc}
0 & 1 \\ 
1 & 0%
\end{array}%
\right) $. The eigenvectors of $h_{\mathbf{k}}$ are 
\begin{equation}
\left\vert \psi _{\mathbf{k}}^{\pm }\right\rangle =\frac{1}{\sqrt{2}}\left( 
\begin{array}{c}
\pm e^{i\theta _{\mathbf{k}}} \\ 
1%
\end{array}
\right) ,
\end{equation}%
with $\theta _{\mathbf{k}}=\arg \left( \vartheta _{\mathbf{k}}\right) $ and
the corresponding eigenvalues 
\begin{equation}
E_{\mathbf{k}}^{\pm }=6\cos k_{y}\pm \left\vert 3+2\cos k_{y}\right\vert
\Lambda _{\mathbf{k}}.
\end{equation}%
Here factor $\Lambda _{\mathbf{k}}=\sqrt{\lambda ^{2}+1+2\lambda \cos k_{x}}$
determines the pseudo band gap, which vanishes at $k_{x}=0,\pi $\ when
taking $\lambda =\pm 1$.

\begin{figure*}[tbh]
	\centering
	\includegraphics[width=1\textwidth]{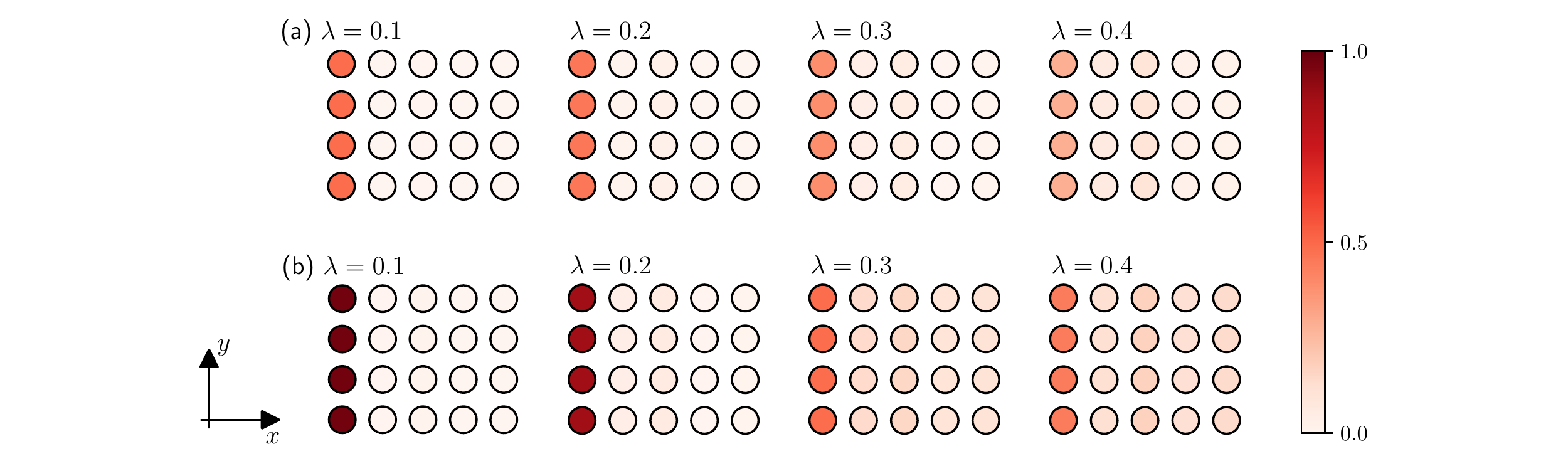}
	\caption{Numerical results for local particle density $\protect\rho \left( 
		\mathbf{r}\right)$ defined in Eq. (\protect\ref{local_density}) for the
		systems with different numbers of fermions and parameters $\protect\lambda$
		. The lattices size and Hubbard interaction strength are all $%
		(N_{x},N_{y})=(5,4)$ and $U=1.5$. The numbers of fermions are two ($%
		\uparrow\downarrow$) and four ($\uparrow\uparrow\downarrow\downarrow$) for
		(a) and (b), respectively.}
	\label{figS1}
\end{figure*}

Under cylindrical boundary condition (taking open boundary condition in $x$
direction), this model supports topological edge states protected by time
reversal symmetry and inversion symmetry. We introduce the wave polarization
to characterize the topological phase transition \cite{liu2017novel,
wu2020nontrivial} 
\begin{equation}
\mathbf{P}^{\pm }=\frac{1}{\left( 2\pi \right) ^{2}}\iint_{\mathrm{BZ}} 
\mathbf{A}_{\mathbf{k}}^{\pm }dk_{x}dk_{y},
\end{equation}%
where $\mathbf{A}_{\mathbf{k}}^{\pm }=\langle \psi _{\mathbf{k}}^{\pm
}|i\partial _{\mathbf{k}}|\psi _{\mathbf{k}}^{\pm }\rangle $\ is the Berry
connection, and the integral region is in the first Brillouin zone. We
simply have%
\begin{equation}
\mathbf{A}_{\mathbf{k}}^{\pm }=-\frac{1}{2}\hat{x}\partial _{k_{x}}\theta _{ 
\mathbf{k}},
\end{equation}%
and 
\begin{equation}
\mathbf{P}^{\pm }=-\frac{1}{2}\hat{x}\mathcal{W}=\frac{1}{2}\hat{x}\left\{ 
\begin{array}{cc}
1, & \left\vert \lambda \right\vert <1 \\ 
0, & \left\vert \lambda \right\vert >1%
\end{array}
\right. .
\end{equation}%
The topological characterization is obvious since 
\begin{equation}
\mathcal{W}=\frac{1}{2\pi }\int_{-\pi }^{\pi }\partial _{k_{x}}\arg \left(
\vartheta _{\mathbf{k}}\right) dk_{x}
\end{equation}%
is essentially the winding number.

In the topologically nontrivial phase $\left\vert \lambda \right\vert <1$,
we have nonzero polarization $\mathbf{P}^{\pm }=\left( 1/2,0\right) $, thus
it is expected to observe the topological edge state in the boundary of $x$
direction when the cylindrical boundary condition is taken. In fact, it can
be checked that in the large-$N_{x}$ limit, the system supports two
degenerate edge modes 
\begin{eqnarray}
|\psi _{k_{y},\sigma }^{\mathrm{L}}\rangle &=&\Omega \sum_{x}\left( -\lambda
\right) ^{x-1}a_{x,k_{y},\sigma }^{\dag }|\mathrm{Vac}\rangle ,  \notag \\
|\psi _{k_{y},\sigma }^{\mathrm{R}}\rangle &=&\Omega \sum_{x}\left( -\lambda
\right) ^{N_{x}-x}b_{x,k_{y},\sigma }^{\dag }|\mathrm{Vac}\rangle ,
\end{eqnarray}%
with the eigenenergy $E^{\mathrm{L}/\mathrm{R}}=6\cos k_{y}$ and the
normalization constant $\Omega =\sqrt{1-\lambda ^{2}}$, where the inverse
transformation is $\left( a_{x,k_{y},\sigma },b_{x,k_{y},\sigma }\right)
=N_{y}^{-1/2}\sum_{y}\left( c_{2x-1,y,\sigma },c_{2x,y,\sigma }\right)
e^{-ik_{y}y}$.

\subsection{More exact numerical results and approximate numerical results}

\label{C} \setcounter{equation}{0} \renewcommand{\theequation}{C%
\arabic{equation}}

\begin{figure*}[tbh]
\centering
\includegraphics[width=1\textwidth]{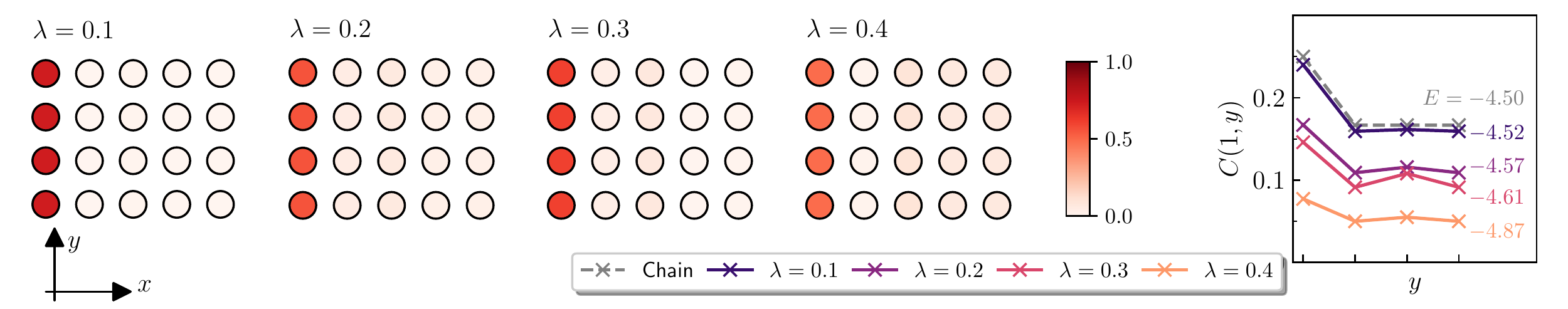}
\caption{Numerical results of the local particle density $\protect\rho %
\left( \mathbf{r}\right)$ and correlator $C\left( 1,y\right) $ for the
system with $3$ ($\uparrow\uparrow\downarrow$) fermions obtained from exact
diagonalization. The lattices size and Hubbard interaction strength are all $%
(N_{x},N_{y})=(5,4)$ and $U=1.5$.}
\label{figS2}
\end{figure*}
\begin{figure}[tbh]
\centering
\includegraphics[width=0.5\textwidth]{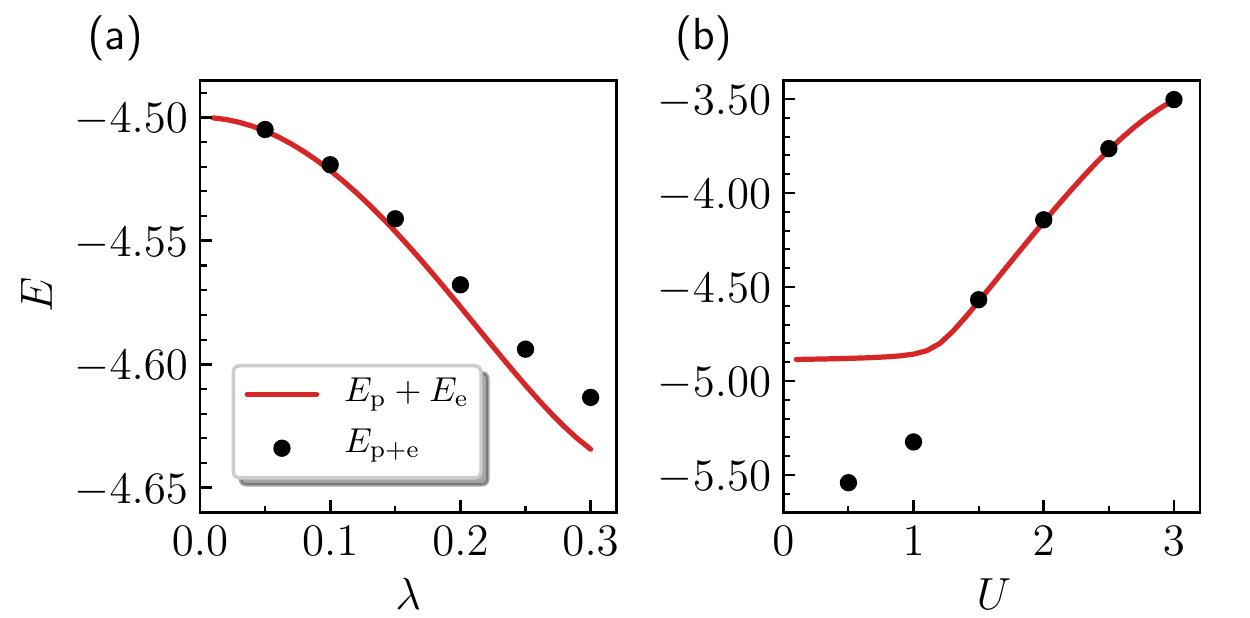}
\caption{Comparison of energy $E_{\mathrm{p+e}}$ with $E_{\mathrm{p}}+E_{ 
\mathrm{e}}$ obtained form exact diagonalization (a) as function of $\protect%
\lambda$ with fixed $U=1.5$; (b) as function of $U$ with fixed $\protect%
\lambda=0.2$. The lattices size is $(N_{x},N_{y})=(5,4)$.}
\label{figS3}
\end{figure}
\begin{figure*}[tbh]
\centering
\includegraphics[width=1\textwidth]{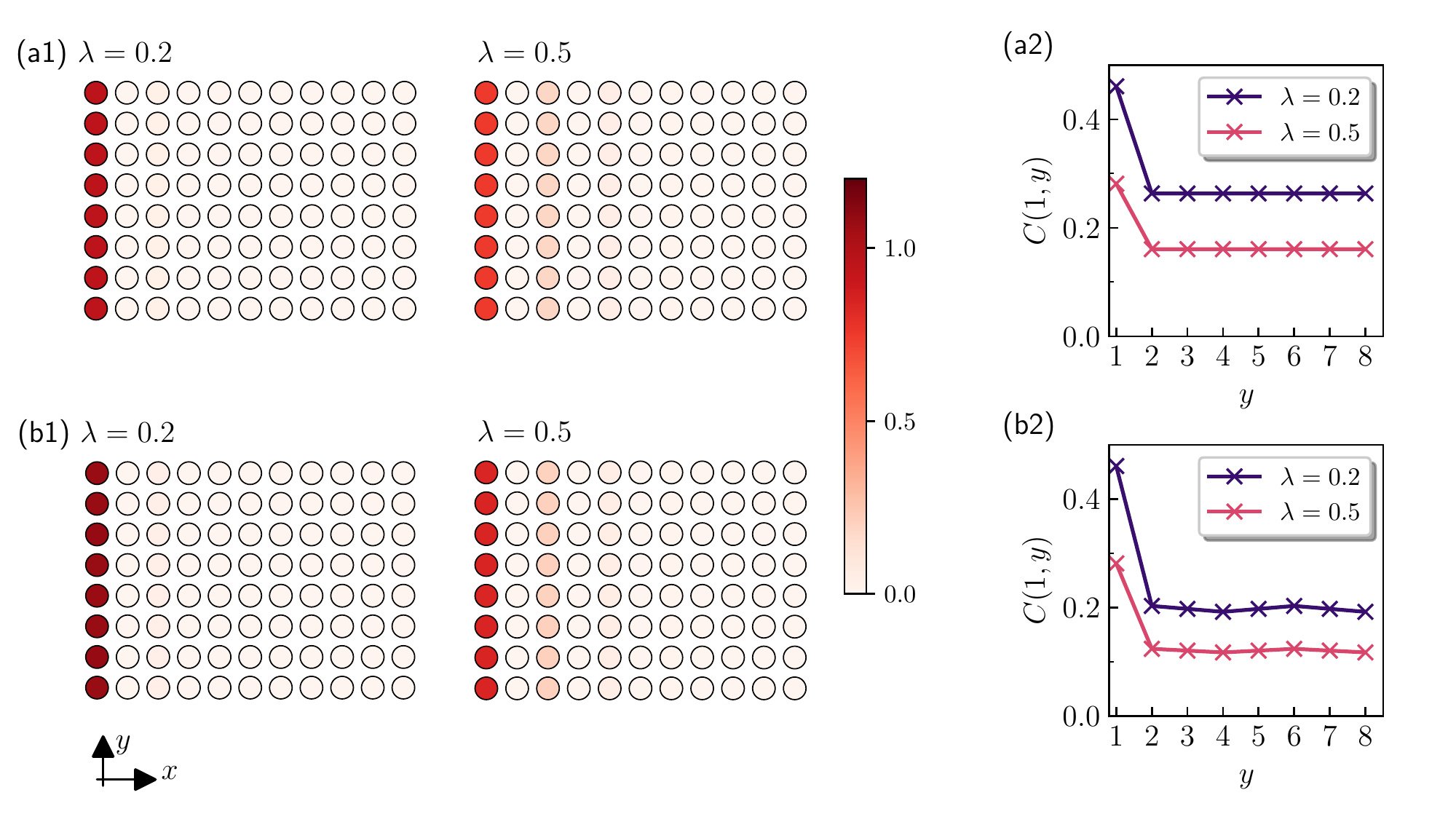}
\caption{Numerical results obtained from the effective Hamiltonian in the
subspace of edge states. (a1) and (b1) are local particle density $\protect%
\rho \left( \mathbf{r}\right)$. (a2) and (b2) show correlator $C\left(
1,y\right) $ for the systems with different numbers of fermions and
parameters $\protect\lambda$. The lattices size and Hubbard interaction
strength are all $(N_{x},N_{y})=(11,8)$ and $U=1.5$. The numbers of fermions
are $8$ ($4\uparrow 4\downarrow$) and $9$ ($5\uparrow 4\downarrow$ ); the
energies are $E=6.00$ and $E=10.24$, for (a) and (b), respectively.}
\label{figS4}
\end{figure*}

To visualize the distribution of the fermions of a state in real space, we
calculate the local particle density 
\begin{equation}
\rho \left( \mathbf{r}\right) =\sum_{\sigma =\uparrow ,\downarrow }\langle
\psi _{\text{\textrm{edge}}}|n_{\mathbf{r},\sigma }|\psi _{\text{\textrm{\
edge }}}\rangle .  \label{local_density}
\end{equation}%
The numerical results of the local particle density $\rho \left( \mathbf{r}%
\right) $ for the $\eta $-pairing edge states with two and four fermions are
presented in Figs. \ref{figS1}(a) and \ref{figS1}(b), which respectively correspond to
the cases considered in Figs. \ref{fig2}(a) and \ref{fig2}(b) of the main text. In Fig. \ref%
{figS2}, we present the numerical results of the three-fermion ($\uparrow
\uparrow \downarrow $) case. It indicates that the three-fermion $\eta $%
-pairing edge state also has off-diagonal long-range order. This is the
simplest two-fluid edge state. We can gain some intuition in term of energy.
We choose one of the three-fermion edge states with energy $E_{\mathrm{p+e}}$
from the numerical data. In Fig. \ref{figS3}, we present the comparison
between the energy $E_{\mathrm{p+e}}$ and $E_{\mathrm{p}}+E_{\mathrm{e}}$,
as functions of $\lambda $ [Fig. \ref{figS3}(a)] and $U$ [Fig. \ref{figS3}%
(b)], where $E_{\mathrm{p}}$ is the energy of $\eta$-pairing edge state in
two-fermion subspace and $E_{\mathrm{e}}$ is the energy of a single-fermion
edge state. It indicates that the energy $E_{\mathrm{p+e}}$ has the form of $%
E_{\mathrm{p}}+E_{\mathrm{e}}$ for small $\lambda $ or large $U$. This
suggests the existence of the two-fluid edge states in certain parameter
region.

To handle a larger system, we can write the matrix of effective Hamiltonian
in the edge-states subspace. For the extended 2D SSH Hubbard model with zero 
$U$, there are two kinds of single-particle states: edge states $|\psi
_{k_{y},\sigma }^{ \mathrm{L/R}}\rangle $ and bulk states $|\psi
_{k_{y},\sigma }^{\mathrm{bulk} }\rangle $, in the topologically nontrivial
phase with $\left\vert \lambda \right\vert <1$. The wave function of an edge
state exponential decay from edge to bulk, then the spatial overlap of these
two states tends to zero in the thermodynamic limit, that is $|\langle \psi
_{k_{y},\sigma }^{\mathrm{\ L/R}}|n_{\mathbf{r} ,\sigma }|\psi
_{k_{y}^{\prime },\sigma }^{\mathrm{bulk} }\rangle |\rightarrow 0$ for any $%
\mathbf{r}$. Since the Hubbard interaction $U\sum_{\mathbf{r}}n_{\mathbf{%
r},\uparrow }n_{\mathbf{r} ,\downarrow }$ is local, when it is switched on,
it does not hybridize these two kinds of states approximately, that is $\sum_{%
\mathbf{r}}|n_{\mathbf{r} ,\sigma }n_{ \mathbf{r},\sigma ^{\prime }}|\psi
_{k_{y},\sigma }^{\mathrm{L/R} }\rangle |\psi _{k_{y}^{\prime },\sigma
^{\prime }}^{\mathrm{bulk}}\rangle |\approx 0 $ $\left( \sigma \neq \sigma
^{\prime }\right) $. The main hybridization occurs between the bulk states,
or the edge states. Notably, this approximation is more effective when $N_{x}
$ is larger. Then we can consider the physics in the invariant subspace of
edge states $\{|\psi _{k_{y},\sigma }^{\mathrm{L/R}}\rangle \}$, which has
bipartite lattice symmetry that ensures the formation of two-fluid edge
states. Based on this approximation, the $\eta$-pairing edge states in a
larger system can be calculated approximately. In Figs. \ref{figS4} (a) and
\ref{figS4}(b), we present the numerical results with $8$ and $9$ fermions on the
lattice with size $(N_{x},N_{y})=(11,8)$, respectively, obtained from the
effective Hamiltonian in the subspace of edge states. We can see that for
the cases of more fermions in a larger system, the $\eta$-pairing and
two-fluid edge states with off-diagonal long-range order still exist.

\subsection{Non-Hermitian description of resonant transmission}

\label{D} \setcounter{equation}{0} \renewcommand{\theequation}{D%
\arabic{equation}}

\begin{figure}[tbh]
\centering
\includegraphics[width=0.5\textwidth]{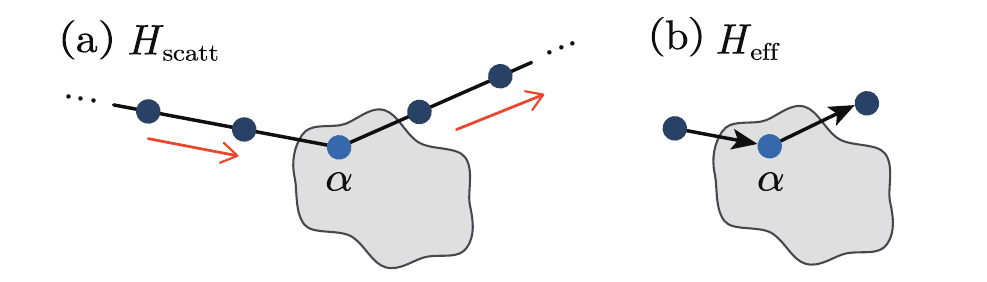}
\caption{(a) Schematic illustration of the Hermitian system $H_{\mathrm{%
scatt }}$ in Eq. (\protect\ref{H_scatt}). The orange arrows represent the
wave of resonant transmission. (b) Schematic illustration of the effective
non-Hermitian system $H_{\mathrm{eff}}$ in Eq. (\protect\ref{H_eff}). The
black arrows indicate the unidirectional hoppings. The gray areas represent
two identical scattering centers $H_{\mathrm{c}}$.}
\label{figS5}
\end{figure}

In this section, we establish a connection between the resonant transmission
of a scattering center and the EP dynamics of a non-Hermitian system, which
is consisted with the scattering center and two extra sites. Two systems are
schematically illustrated in Fig. \ref{figS5}(a) and \ref{figS5}(b). We consider a
general scattering system by a single-particle Hamiltonian 
\begin{figure}[tbh]
\centering
\includegraphics[width=0.5\textwidth]{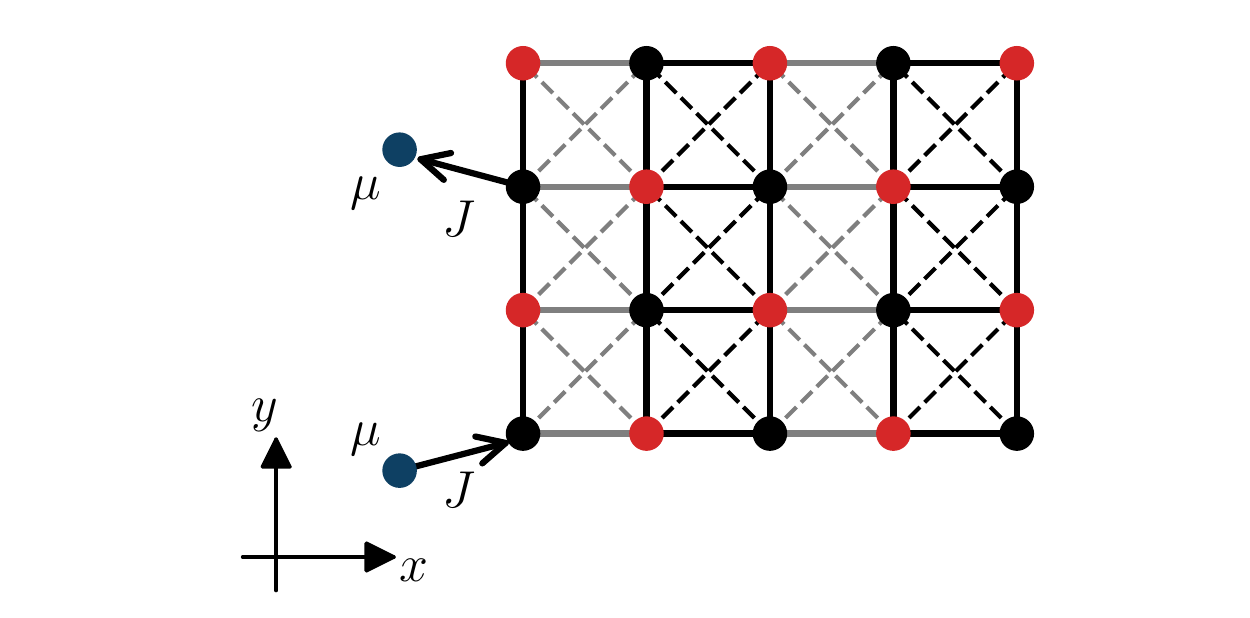}  
\caption{Schematic illustration of the configuration of the non-Hermitian
Hamiltonian $H_{\text{\textrm{NH}}}$ considered in the main text. The blue
dots represent two extra sites with on-site potential of strength $\protect%
\mu$. The arrows indicate the unidirectional hopping of strength $J$.}
\label{figS6}
\end{figure}
\begin{equation}
H_{\mathrm{scatt}}=H_{\mathrm{Ld}}+H_{\mathrm{c}},  \label{H_scatt}
\end{equation}%
where $H_{\mathrm{Ld}}$ represents the two\ leads,%
\begin{eqnarray}
H_{\mathrm{Ld}} &=&J\sum_{j=1}^{\infty }\left( \left\vert -j\right\rangle
\left\langle -j-1\right\vert +\left\vert j\right\rangle \left\langle
j+1\right\vert \right) \\
&&+J\left( \left\vert -1\right\rangle \left\langle \alpha \right\vert
+\left\vert 1\right\rangle \left\langle \alpha \right\vert \right) +\mathrm{%
\ H.c.}+\mu \sum_{\left\vert j\right\vert =1}^{\infty }\left\vert
j\right\rangle \left\langle j\right\vert ,  \notag
\end{eqnarray}%
while $H_{\mathrm{c}}$\ is a scattering center of $N$ sites,%
\begin{equation}
H_{\mathrm{c}}=\sum_{q=1}^{N}\varepsilon _{q}\left\vert \phi
_{q}\right\rangle \left\langle \phi _{q}\right\vert .
\end{equation}%
Here $\alpha $ represents a site in the cluster $H_{\mathrm{c}}$ connecting
to the left and right leads. $\left\vert \phi _{q}\right\rangle $ denotes
the normalized eigenstate of $H_{\mathrm{c}}$\ with energy $\varepsilon _{q}$%
.\ Now we consider the case, in which $H_{\mathrm{c}}$\ has an isolated
energy level at $q=\rho $, satisfying $\mu =\varepsilon _{\rho }$ and $%
\left\vert \varepsilon _{\rho }-\varepsilon _{\rho \pm 1}\right\vert \gg J$.
It can be checked that the state $|\psi _{\pi /2}\rangle $\ in the form 
\begin{equation}
|\psi _{\pi /2}\rangle =\left\{ 
\begin{array}{cc}
e^{-i\pi j/2}\left\vert j\right\rangle , & \left\vert j\right\vert \geqslant
1 \\ 
\gamma \left\vert \phi _{\rho }\right\rangle , & j\in \left\{ \mathrm{c}
\right\}%
\end{array}
\right. ,
\end{equation}%
is an eigenstate of $H_{\mathrm{scatt}}$ with energy $\varepsilon _{\rho }$.
where $\left\{ \mathrm{c}\right\} $\ denotes the set of index for the sites
of scattering center. Here $\gamma $ is a complex number, determined by\ $%
\left\langle \alpha \right\vert \phi _{\rho }\rangle =\gamma ^{-1}$.

In parallel, we consider a non-Hermitian Hamiltonian 
\begin{eqnarray}
H_{\mathrm{eff}} &=&J\left( \left\vert \alpha \right\rangle \left\langle
-1\right\vert +\left\vert 1\right\rangle \left\langle \alpha \right\vert
\right)  \notag \\
&&+\sum_{q=1}^{N}\varepsilon _{q}\left\vert \phi _{q}\right\rangle
\left\langle \phi _{q}\right\vert +\mu \sum_{j=\pm 1}\left\vert
j\right\rangle \left\langle j\right\vert ,  \label{H_eff}
\end{eqnarray}%
which contains two unidirectional hopping terms. For the isolated energy
level with $\left\vert \varepsilon _{\rho }-\varepsilon _{\rho \pm
1}\right\vert \gg J$, it reduces to%
\begin{eqnarray}
H_{\mathrm{eff}} &\approx &J\left( \left\vert \alpha \right\rangle
\left\langle -1\right\vert +\left\vert 1\right\rangle \left\langle \alpha
\right\vert \right)  \notag \\
&&+\varepsilon _{\rho }\left\vert \phi _{\rho }\right\rangle \left\langle
\phi _{\rho }\right\vert +\mu \sum_{j=\pm 1}\left\vert j\right\rangle
\left\langle j\right\vert .
\end{eqnarray}%
Under the resonant condition $\mu =\varepsilon _{\rho }$\ the dynamics is
governed by the Jordan block with order of three 
\begin{equation}
M_{\mathrm{JB}}=\left( 
\begin{array}{ccc}
\mu & 0 & 0 \\ 
J & \mu & 0 \\ 
0 & J & \mu%
\end{array}%
\right) .
\end{equation}%
For initial state $\left\vert \psi (0)\right\rangle =\left\vert
-1\right\rangle =\left( 1,0,0\right) ^{\mathrm{T}}$, the final state is $%
\left\vert \psi (\infty )\right\rangle =\left\vert 1\right\rangle =\left(
0,0,1\right) ^{\mathrm{T}}$, due to the fact%
\begin{eqnarray}
&&\exp \left( -iM_{\mathrm{JB}}t\right)  \notag \\
&=&\exp \left( -i\mu t\right) \left[ 1-i\left( M_{\mathrm{JB}}-\mu \right) t-%
\frac{1}{2}\left( M_{\mathrm{JB}}-\mu \right) ^{2}t^{2}\right] ,  \notag
\end{eqnarray}%
and 
\begin{equation}
\exp \left( -iM_{\mathrm{JB}}t\right) \left( 
\begin{array}{c}
1 \\ 
0 \\ 
0%
\end{array}%
\right) =\exp \left( -i\mu t\right) \left( 
\begin{array}{c}
1 \\ 
-iJt \\ 
-\frac{1}{2}J^{2}t^{2}%
\end{array}%
\right) .
\end{equation}%
A straightforward implication of the result is that this process accords
with the resonant transmission, i.e., the particle is perfectly transported
from the left to the right.

Nevertheless, we would like to point that no matter if it is resonant or
not, the Jordan block always exists, but with different order. Here we
consider the case of $\mu $ deviate from $\varepsilon _{\rho }$, and the
dynamics is governed by the matrix 
\begin{equation}
M=\left( 
\begin{array}{ccc}
\mu  & 0 & 0 \\ 
J & \varepsilon _{\rho } & 0 \\ 
0 & J & \mu 
\end{array}%
\right) \text{.}
\end{equation}%
Here matrix $M$ can be related to the Jordan block with order of two, which
is $\left( 
\begin{array}{ccc}
\mu  & 0 & 0 \\ 
1 & \mu  & 0 \\ 
0 & 0 & \varepsilon _{\rho }%
\end{array}%
\right) $, by Jordan decomposition. The time evolution is 
\begin{equation}
\exp \left( -iMt\right) \left( 
\begin{array}{c}
1 \\ 
0 \\ 
0%
\end{array}%
\right) =\exp \left( -i\mu t\right) \left( 
\begin{array}{c}
1 \\ 
\varsigma \tau  \\ 
\varsigma ^{2}\left[ \tau +i\left( \varepsilon _{\rho }-\mu \right) t\right] 
\end{array}%
\right) , 
\end{equation}%
where $\varsigma =J/\left( \varepsilon _{\rho }-\mu \right) $ and $\tau
=\exp \left[ -i\left( \varepsilon _{\rho }-\mu \right) t\right] -1$.

It indicates that in the resonant case, the evolved state converges to the
target state more rapidly. In other word, to distinguish two different
processes, we can observe the fidelity between the evolved state and the
target state for two different dynamics processes at the same sufficiently
long time $t$.

In Fig. \ref{figS6}, we present the schematic illustration of the system
configuration of the Hamiltonian in Eq. (13) of the main text.

\end{document}